\documentclass[onecolumn,floatfix,superscriptaddress,a4paper,showpacs,showkeys,nofootinbib,reprint]{revtex4-1}

\usepackage{enumitem}
\usepackage{graphicx}
\usepackage{float}
\usepackage[utf8]{inputenc}
\usepackage{bm}
\usepackage[colorlinks=true, linkcolor=blue,citecolor=blue,allcolors=blue]{hyperref}
\usepackage{tabularx}
\usepackage{xcolor}
\usepackage{xspace}
\usepackage[capitalize]{cleveref}
\usepackage{makecell}
\usepackage{tabularx,booktabs}

\graphicspath{{./figures/}}

\newcommand{\snnh}{${\sqrt{s_{_{NN}}}=2.4}$~GeV\xspace}

\begin{document}

\title{
Ambiguities in the hadro-chemical freeze-out of Au+Au collisions at SIS18 energies and how to resolve them
}
\date{\today}
\author{Anton Motornenko}
\affiliation{
Institut f\"ur Theoretische Physik,
Goethe Universit\"at, Max-von-Laue-Str. 1, D-60438 Frankfurt am Main, Germany}
\affiliation{
Frankfurt Institute for Advanced Studies, Giersch Science Center,
Ruth-Moufang-Str. 1, D-60438 Frankfurt am Main, Germany}
\author{Jan~Steinheimer}
\affiliation{
Frankfurt Institute for Advanced Studies, Giersch Science Center,
Ruth-Moufang-Str. 1, D-60438 Frankfurt am Main, Germany}
\author{Volodymyr Vovchenko}
\affiliation{Nuclear Science Division, Lawrence Berkeley National Laboratory, 1 Cyclotron Road, Berkeley, CA 94720, USA}
\author{Reinhard Stock}
\affiliation{
Frankfurt Institute for Advanced Studies, Giersch Science Center,
Ruth-Moufang-Str. 1, D-60438 Frankfurt am Main, Germany}
\affiliation{
Institut f\"ur Kernphysik,
Goethe Universit\"at Frankfurt, Max-von-Laue-Str. 1, D-60438 Frankfurt am Main, Germany}
\author{Horst Stoecker}
\affiliation{
Institut f\"ur Theoretische Physik,
Goethe Universit\"at, Max-von-Laue-Str. 1, D-60438 Frankfurt am Main, Germany}
\affiliation{
Frankfurt Institute for Advanced Studies, Giersch Science Center,
Ruth-Moufang-Str. 1, D-60438 Frankfurt am Main, Germany}
\affiliation{
GSI Helmholtzzentrum f\"ur Schwerionenforschung GmbH, D-64291 Darmstadt, Germany}

\begin{abstract}
    The thermal fit to preliminary HADES data of Au+Au collisions at $\sqrt{s_{_{NN}}}=2.4$ GeV shows two degenerate solutions at $T\approx50$ MeV and $T\approx70$ MeV. The analysis of the same particle yields in a transport simulation of the UrQMD model yields the same features, i.e. two distinct temperatures for the chemical freeze-out. While both solutions yield the same number of hadrons after resonance decays, the feeddown contribution is very different for both cases. This highlights that two systems with different chemical composition can yield the same multiplicities after resonance decays.
    The nature of these two minima is further investigated by studying the time-dependent particle yields and extracted thermodynamic properties of the UrQMD model. It is confirmed, that the evolution of the high temperature solution resembles cooling and expansion of a hot and dense fireball. The low temperature solution displays an unphysical evolution: heating and compression of matter with a decrease of entropy. These results imply that the thermal model analysis of systems produced in low energy nuclear collisions is ambiguous but can be interpreted by taking also the time evolution and resonance contributions into account.  
\end{abstract}

\maketitle

\section{Introduction}

A recent statistical model analysis~\cite{Harabasz:2020sei} of the new HADES collaboration data yields an unexpectedly low temperature best $\chi^2$ fit for hadron yields in a standard chemical freeze-out model.
Statistical models are a well established tool to describe hadron production in heavy ion collisions in experiments at the Bevalac,  SIS, AGS, SPS, RHIC, and LHC accelerators. It is of great interest that static statistical models with only a handful of thermodynamic parameters provide a surprisingly good description of the particle yields from system with complicated non-equilibrium dynamics and interactions.
The ideal hadron resonance gas model~(HRG) gives a generally good description of the many experimentally observed hadron yields measured at various collision energies~\cite{Cleymans:1992zc,BraunMunzinger:1996mq,Becattini:2000jw,Becattini:2003wp,Andronic:2017pug}. This approach assumes that the chemical composition of the system, and thus the final particle multiplicities, are fixed at late stage of heavy ion collisions, the so-called chemical freeze-out.
The chemical freeze-out parameters for each collision system are obtained through a fit to measured particle multiplicities.
Collision energy dependence of the extracted parameters like temperature and chemical potentials defines the chemical freeze-out line, mapping heavy-ion collision experiments to the QCD phase diagram~\cite{Cleymans:2005xv,Andronic:2005yp}.
Although the statistical model has different implementations, which can mainly vary in the way hadronic interactions are treated~\cite{Zschiesche:2002zr,Satarov:2016peb,Vovchenko:2018fmh,Andronic:2018qqt,Poberezhnyuk:2019pxs}, with few exceptions~\cite{Vovchenko:2015cbk}, they give very similar chemical freeze-out curves.
Consequently, the existence of a universal description of the last point of chemical equilibrium in heavy ion collisions is of great importance to the interpretation of heavy ion experimental data.
Any deviations or discrepancies with this picture would therefore be of great interest, having strong implications on the applicability of the standard paradigm governing high energy heavy ion collisions. 

This paper shows that the hadron yields in the few GeV collision energy regime pose a conundrum for the thermal model fit. The particle yields measured by the HADES collaboration at SIS18 accelerator at GSI can be described similarly well by two distinct sets of thermodynamic parameters. 
The same degeneracy is observed in transport simulations within the UrQMD model.
Studying the time evolution of the system within UrQMD allows to elaborate on the viability of the two solutions.

\begin{figure*}[t]
    \centering
    \includegraphics[width=0.49\textwidth]{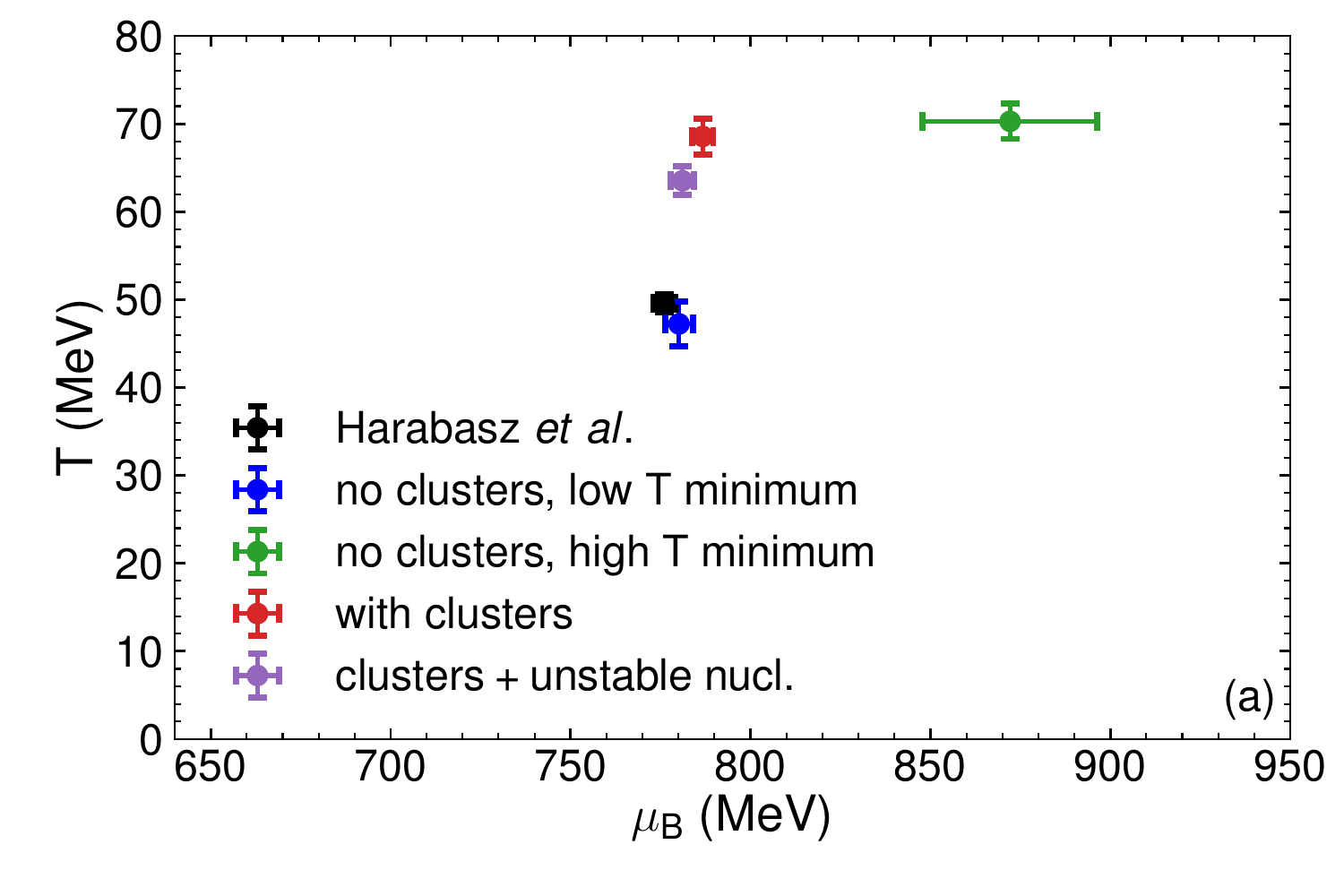}
    \includegraphics[width=0.49\textwidth]{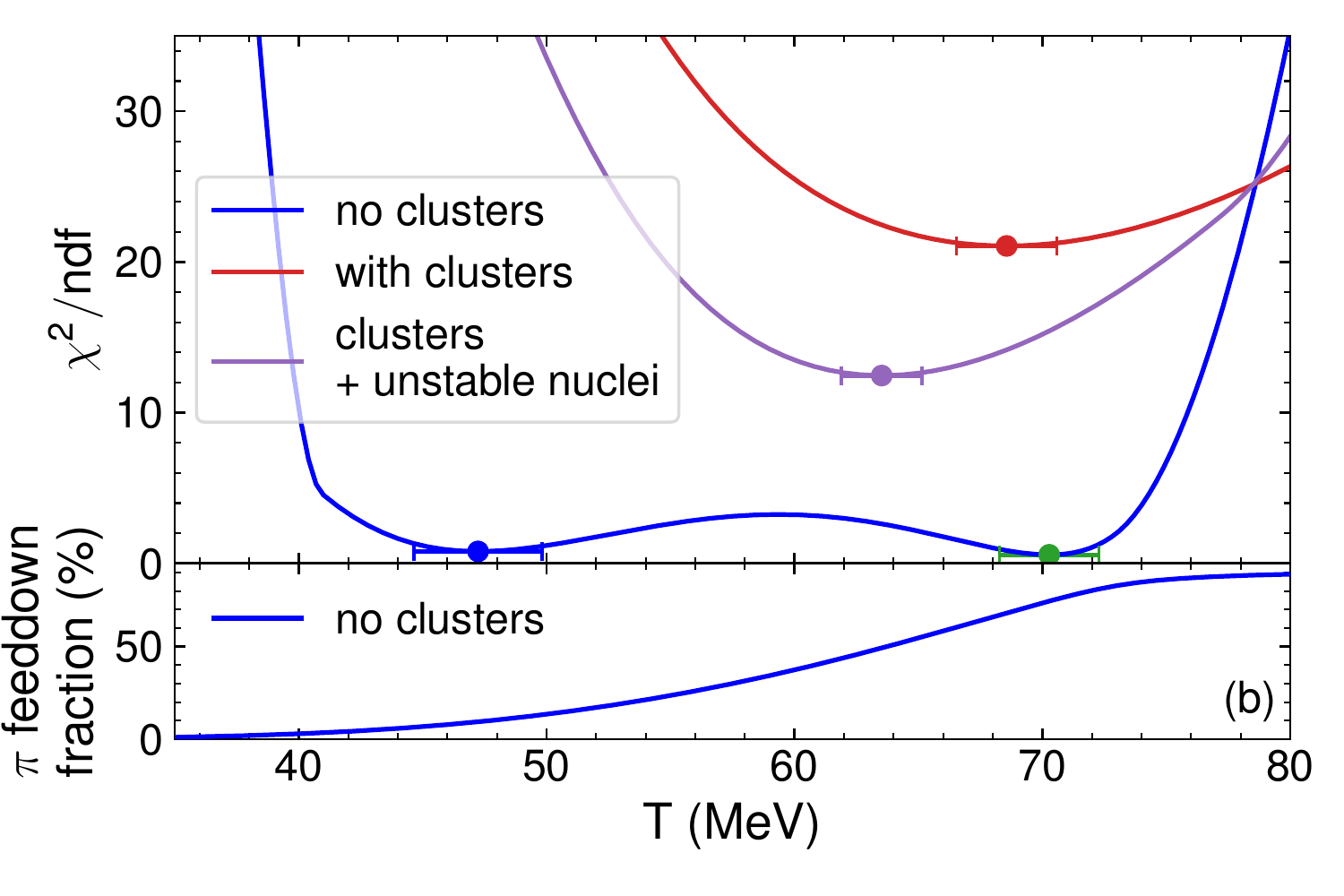}
    \caption{{\bf (a):} Chemical freeze-out parameters in $\mu_B$-$T$ plane obtained from thermal model analysis of the hadron yields measured by the HADES collaboration, Table~\ref{tab:mult}. Different locations of the freeze-out correspond to different considered freeze-out scenarios, see text for details. {\bf (b):} $\chi^2$ profiles of the performed fits. The ``no clusters'' scenario gives the best fit, however produces two distinct minima. 
    To distinguish between the two minima the $\pi$ feeddown fraction is presented which is calculated as a ratio of the number of charged pions that stem from resonance decays to the total number of charged pions, $({\pi^+ + \pi^-})_{\rm feeddown}/({\pi^+ + \pi^-})_{\rm total}$.}
    \label{fig:muT-chi2}
\end{figure*}

\section{Dataset used}
The HADES collaboration at the SIS18 accelerator at GIS has measured a set of preliminary hadron multiplicities at 10\% most central Au+Au collisions at \snnh~\cite{Szala:2019,Adamczewski-Musch:2017rtf,Adamczewski-Musch:2018xwg,Adamczewski-Musch:2020vrg}. These data are summarized in Table~\ref{tab:mult}.

\begin{table}[b]
\begin{center} 
\begin{tabular}{cccc}
\hline \hline
    $\text{particle}$  & $\text{multiplicity}$ \ \ \  & $\text{uncertainty}$ & $\text{Ref.}$ \\ \hline 
    $p$      & 77.6   & $\pm 2.4$ & \text{\cite{Szala:2019}}\\
     $p + n \rightarrow {^{2}H}$ & 28.7 & $\pm 0.8$ & \text{\cite{Szala:2019}}\\
     $p + 2n \rightarrow {^{3}H}$ & 8.7   & $\pm  1.1$ & \text{\cite{Szala:2019}}\\
     $p + p + n \rightarrow {^{3}He}$ & 4.6  & $\pm  0.3$ & \text{\cite{Szala:2019}} \\
    $p \: \text{(bound)}$ & 46.5 &  $\pm  1.5$ &\text{\cite{Szala:2019}}\\
    $\pi^+$  & 9.3   &  $\pm 0.6$ & \text{\cite{Adamczewski-Musch:2020vrg}}\\
    $\pi^-$  & 17.1   & $\pm 1.1$ & \text{\cite{Adamczewski-Musch:2020vrg}}\\
    $K^+$    & $5.98\,10^{-2}$ & $\pm 6.79\,10^{-3}$  & \text{\cite{Adamczewski-Musch:2017rtf}} \\
    $K^-$    & $5.6\,10^{-4}$ & $\pm 5.96 \,10^{-5}$ & \text{\cite{Adamczewski-Musch:2017rtf}} \\
    $\Lambda$ & $8.22\,10^{-2}$ & $^{+5.2} _{-9.2}\,10^{-3}$ & \cite{Adamczewski-Musch:2018xwg} \\ 
    \hline \hline
\end{tabular}

\caption{Preliminary particle yields measured by the HADES collaboration at SIS18 accelerator, \snnh, for 10\% most central Au-Au collisions. Protons bound in nuclei can be accounted all as free under the assumption that the nuclei are formed after kinetic freeze-out. The data compilation is extracted from~\cite{Harabasz:2020sei}.
\label{tab:mult}}
\end{center}
\end{table}

A thermal model analysis of these preliminary HADES data~\ref{tab:mult} has been previously published in Ref.~\cite{Harabasz:2020sei}. 
The authors extracted thermodynamic properties of the system created in the heavy ion collisions by assuming that all hadron multiplicities are fixed at a single chemical freeze-out.
The freeze-out of the system was described within an ideal HRG model. 
Additionally, it was assumed that light nuclei are not present at the chemical freeze-out stage, but that they are formed instead at a later stage via a different mechanism.
The protons calculated at the chemical freeze-out do therefore include both, the protons  measured as free protons and those which are later bound into light nuclei.
The analysis used
$p$, $\pi^{+,-}$, ${\rm K}^{+,-}$, and $\Lambda$ yields as input to the fit. 
The measured proton yield was calculated as a sum of the directly measured unbound protons and protons bound inside the measured light nuclei.
Here the same dataset is used as in Ref.~\cite{Harabasz:2020sei}.

\section{Thermal model analysis}

\begin{table*}[t]
    \centering
    \begin{tabular}{c|c|c|c|c|c}
    \hline
    \hline
    Parameter & \makecell{Harabasz {\it et al.}~\cite{Harabasz:2020sei}} & \makecell{no clusters\\ low $T$ minimum} & \makecell{no clusters\\ high $T$ minimum} &  with clusters & \makecell{with clusters\\ + unstable nuclei}\\
    \hline
    $T$ (MeV) & $49.6 \pm 1.1$ & $47.2 \pm 2.6$ & $70.3 \pm 2.0$ & $68.6 \pm 2.0$ & $63.5 \pm 1.6$ \\
    $R$ (fm) & $ 16.0 $ & $18.9 \pm 2.2$ & $6.8 \pm 0.9$ & $9.0 \pm 0.4$ & $10.4 \pm 0.3$ \\
    $\mu_B$ (MeV) & $776 \pm 3$ & $780.1 \pm 3.8$ & $872.1 \pm 24.3$ & $786.7 \pm 2.9$ & $781.1 \pm 3.3$ \\
    $\gamma_S$ & $0.16 \pm 0.02$ & $0.19 \pm 0.07$ & $0.05 \pm 0.01$ & $0.03 \pm 0.01$ & $0.04 \pm 0.01$ \\
    $\chi^2/N_{\rm df}$ & $ N_{\rm df}=0 $ & $1.58/2$ & $1.13/2$ & $105.30/5$ & $62.30/5$ \\
    \hline
    \hline
    \end{tabular}
    \caption{Summary of the fitted parameters in $\sqrt{s_{_{NN}}}=2.4$ GeV Au+Au collisions. These results are illustrated in Fig.~\ref{fig:muT-chi2}.}
    \label{tab:params}
\end{table*}

The thermal model analysis performed in Ref.~\cite{Harabasz:2020sei} suggested that the chemical freeze-out in Au+Au collisions at \snnh occurs at  rather cold and dilute system, characterized by the following set of thermodynamical parameters: $T$=~49.6~$\pm$~1~MeV, $\mu_B$=~776~$\pm$~3~MeV, $\mu_{I_3}=-14.1 \pm 0.2$~MeV, $\mu_S$=~123.4~$\pm$~2~MeV, and $\gamma_s$=~0.16$\pm$~0.02. 
In the following first a similar analysis using the same data is performed.
Here it is assumed that the particle yields are fixed at the chemical freeze-out which is described by ideal HRG in the grand canonical ensemble. The freeze-out state is then related to the thermodynamic parameters of a hadron resonance gas: temperature $T$, baryon, electric, and strangeness chemical potentials $\mu_B$, $\mu_Q$, and $\mu_S$, respectively. 
The system size is defined through the freeze-out radius $R$ or volume $V=\frac{4}{3}\pi R^3$. 
Strangeness undersaturation is taken into account via a strangeness suppression factor~$\gamma_S$~\cite{Letessier:2005qe}. 
The electric and strangeness chemical potentials $\mu_Q$ and $\mu_S$ are fixed by the charge (electric and strange) content of the colliding nuclei, namely the total strangeness vanishes $n_S=0$ and the electric to baryon charge ratio equals to $n_Q/n_B=0.4$. It should be noted that in the Ref.~\cite{Harabasz:2020sei} $\mu_Q$ and $\mu_S$ were used as free fit parameters, this led to $N_{\rm df}=0$ and resulted in a slightly different set of thermodynamical parameters, as compared to the analysis presented here.

The ideal HRG model is used, i.e. at the freeze-out the system is represented by a multi-component gas of free hadrons and resonances in equilibrium. 
The particle list consists of hadrons listed in Particle Data Tables 2020~\cite{Zyla:2020zbs} with an \textit{established} status.
Our analysis is performed using the open source \texttt{Thermal-FIST} package version 1.3~\cite{Vovchenko:2019pjl}. 

In an alternative scenario one can assume that the yields of light nuclei are fixed at the chemical freeze-out together with all other hadrons.
The thermal model works remarkably well for describing the yields of light nuclei across a broad range of collision energies~\cite{Hahn:1986mb,BraunMunzinger:1994iq,Andronic:2010qu}.
Thus, in both the discussed scenarios, the formation of nuclei at a late stage due to coalescence from primordial nucleons as well the direct creation of nuclei at the chemical freeze-out can give similar results on the yields of nuclei. 
It is challenging to explain how the nuclei may survive the temperatures which are an order higher than their binding energies if the second scenario is the correct one.
However, some progress on the understanding of this phenomenon has recently been made~\cite{Oliinychenko:2018ugs,Xu:2018jff,Vovchenko:2019aoz}.
See also~\cite{Mrowczynski:2016xqm,Bellini:2018epz,Mrowczynski:2020ugu} for possible ways to distinguish between thermal model and coalescence.

To take into account the different possibilities for the mechanism of light nuclei production, the experimental data on particle yields is analyzed in three different setups:
\begin{enumerate}[label=(\alph*)]
    \item {\bf no clusters} -- assumes that light nuclei are formed after the chemical freeze-out, thus light nuclei are to be omitted from the thermal model particle list.
    Those protons that later-on bind into the light nuclei are counted as 'free' protons at the chemical freeze-out.
    This is the scenario that had been considered in Ref.~\cite{Harabasz:2020sei}.
    \item {\bf clusters included} -- light nuclei are formed at the chemical freeze-out.
    Only stable light nuclei are included in the thermal model particle list.
    \item {\bf clusters and decays of unstable nuclei are included} -- 
    the thermal model additionally includes the feeddown contributions from the decays of unstable $A=4$, and $A=5$ nuclei to the final yields of protons, deuterons, tritons, $^3$He, and $^4$He at the chemical freeze-out, as discussed in~\cite{Vovchenko:2020dmv}.
\end{enumerate}

\begin{figure}[h!]
    \centering
    \includegraphics[width=0.49\textwidth]{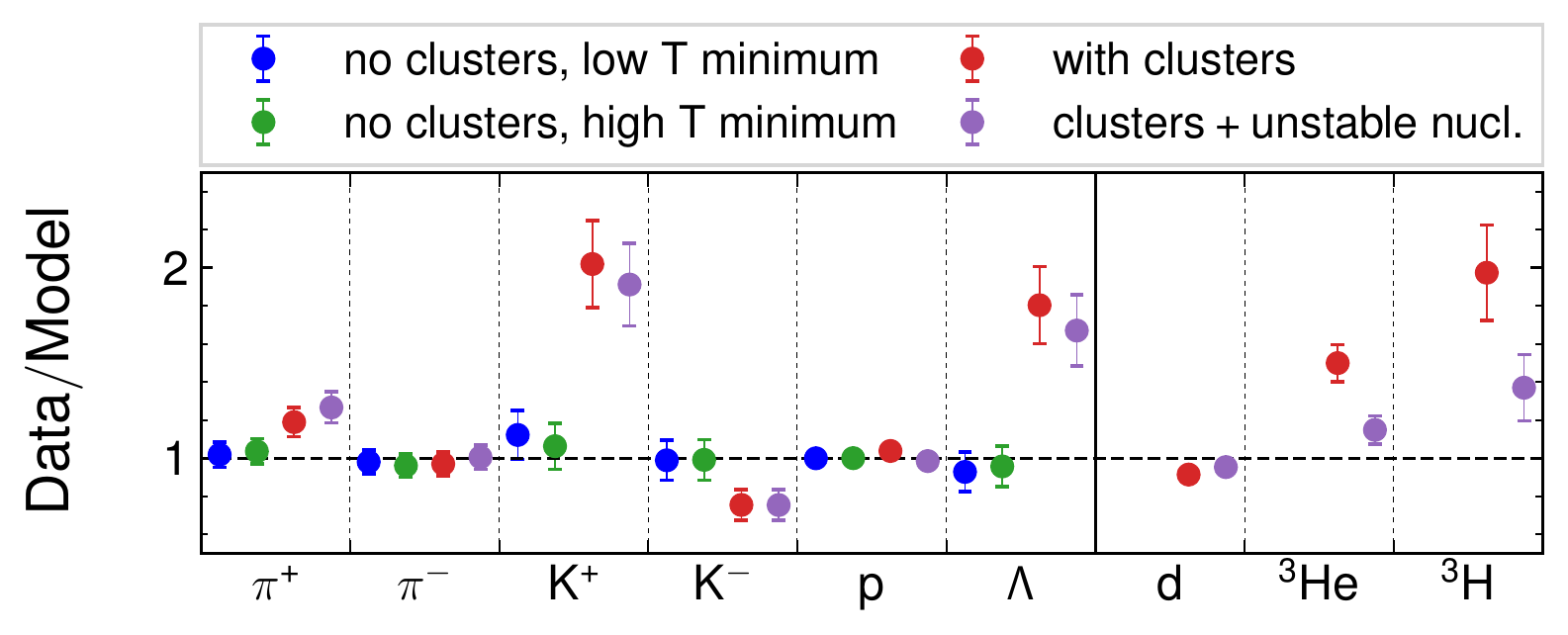}
    \caption{The data-to-model ratios as resulting from the distinct three scenarios of thermal fits
    to the measured particle yields in the 10\% most central Au-Au collisions at \snnh.}
    \label{fig:model_data}
\end{figure}

The resulting thermodynamic parameters obtained within these three scenarios are presented in Fig.~\ref{fig:muT-chi2} and Table~\ref{tab:params} 
and compared with the analysis of the data from Ref.~\cite{Harabasz:2020sei}.
The corresponding data/model ratios for the fitted yields are presented in Fig.~\ref{fig:model_data}.

The first scenario, where light nuclei are not included in the particle list, shows a similar fit as the one in Ref.~\cite{Harabasz:2020sei}. 
However, a closer look at the resulting $\chi^2$ profiles
in Fig.~\ref{fig:muT-chi2} reveals that this scenario actually has two degenerate solutions. 
The data can be described by two distinct sets of thermodynamic parameters with similarly good accuracy. 
The $\chi^2$ profile as a function of the freeze-out temperature has two minima located at $T\approx 50$ MeV and $T\approx70$ MeV 
These minima are referred as ``low temperature'' and ``high temperature'' minimum, respectively. 
Both these two minima describe the data well, characterized by $\chi^2/N_\mathrm{df} < 1$.
Only the $T \approx 50$~MeV minimum was discussed in~\cite{Harabasz:2020sei}~\footnote{We were able to reproduce the values obtained in~\cite{Harabasz:2020sei},
in this case the second minimum was also present.}.

To elaborate on the differences between the two minima a detailed look at the role of resonance feeddown is taken.
The bottom panel of Fig.~\ref{fig:muT-chi2} (b) shows the 
fraction of charged pions coming from resonance decays as a function of the temperature for the ``no clusters'' scenario.
It is clear that the feeddown fraction is significantly larger at the high-temperature minimum. This implies that the two minima give similar final yields of measured particles but correspond to significantly different chemical composition of the primordial hadrons.
At the high temperature more hadrons are present in the form of excited states while the low temperature system corresponds to essentially a gas of ground state hadrons.\footnote{Note, this case of double minima is rather different from a similar finding from the analysis of the ALICE data within excluded volume HRG model in~\cite{Vovchenko:2015cbk} where the high-temperature minimum was connected with the strong effect of excluded-volume interactions~\cite{Vovchenko:2016ebv,Satarov:2016peb}.}
Thus, determining the feeddown fractions is one possible way to distinguish the two minima.
This can potentially be achieved via a statistical model analysis of yields of short-lived resonances like $\rho^0$ or ${\rm K}^*$. Such an analysis may require incorporating partial chemical equilibrium into the HRG model, as discussed in~\cite{Motornenko:2019jha}. The yield of the unstable $\phi$ meson is already measured by the HADES collaboration~\cite{Adamczewski-Musch:2017rtf}. This data may be used in the mentioned partial chemical equilibrium approach, however a careful treatment of the mesons hidden strangeness should be carried out. A strangeness-canonical approach was found to describe well the $\phi$ meson yield in lighter systems~\cite{Agakishiev:2015bwu}, there a fit to $\sqrt{s_{_{NN}}}=2.61$ GeV Ar+KCl data resulted in the freeze-out temperature of $T=70\pm3$ MeV. Recently it was also pointed out that a thermal model analysis of the unstable $\Delta$-baryon yield in Au+Au collision at the SIS18 energies provides $T\approx 70$ MeV which is close to the high temperature minimum~\cite{Reichert:2020uxs}.

Another difference between the two solutions lies in their thermodynamical properties: the high temperature minimum corresponds to the freeze-out baryon density of $n_B^{{\rm high}T}\approx0.22$~fm$^{-3}$, while for the low temperature minimum it is much lower at $n_B^{{\rm low}T}\approx0.01$~fm$^{-3}$.
The high temperature description of the fireball created at SIS18 energies may also improve the Siemens-Rasmussen description of the rapidity distribution presented in~\cite{Harabasz:2020sei}, where the calculated distributions were found to be too narrow as compared to experimentally measured $dN/dy$.

\begin{figure*}[t]
    \centering
    \includegraphics[width=0.7\textwidth]{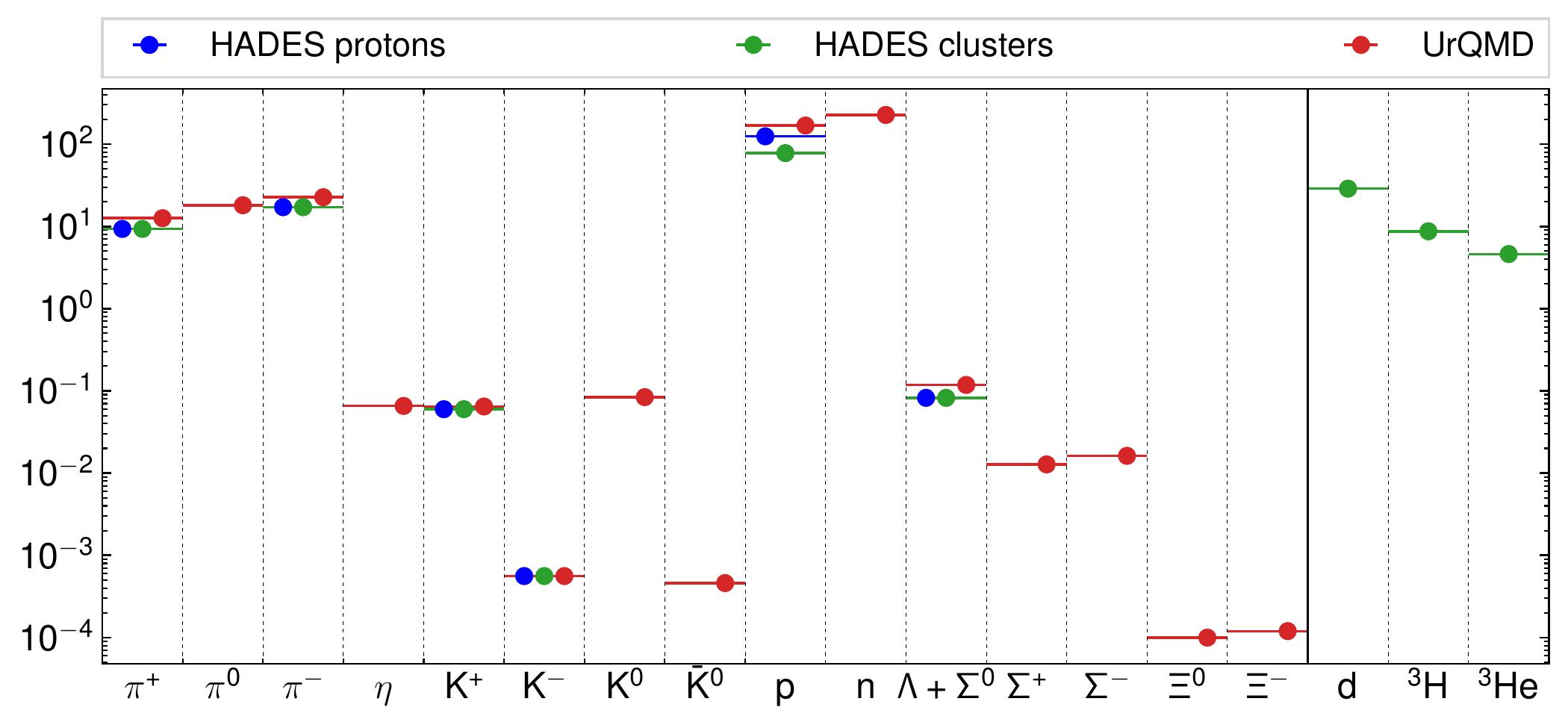}
    \caption{4$\pi$ particle yields in Au+Au collisions at \snnh measured by the HADES collaboration and predicted by the UrQMD model. The experimental data is presented for all measured particle yields including light nuclei (HADES clusters), and for the case when protons bound in light nuclei are added to the number of free protons (HADES protons). 
    The UrQMD data correspond to particle yields calculated in $4\pi$ 10\% most central Au+Au events at $\sqrt{s_{_{NN}}}=2.4$~GeV.
    The error bars of the presented particle yields are smaller than the symbol size.
    }
    \label{fig:yields_urqmd}
\end{figure*}

Both minima describe every measured hadron yield remarkably well, thus, the presently available data alone does not allow to favor one set of parameters over another. Additional information is required.
In Sec.~\ref{sec:UrQMD} this question is discussed by utilizing transport model simulations of Au-Au collisions at SIS18 energies.

Next,  the possibility that the yields of light nuclei are fixed at the chemical freeze-out is investigated.
For this scenario the measured yields of light nuclei are included in the fit and, therefore, the yields of bound protons are excluded from the free protons yield at the chemical freeze-out. 
Here the fits reveal only a single $\chi^2$ minimum. 
However, the fit quality worsens significantly, giving $\chi^2/\mathrm{ndf} \gtrsim 20$.

As was discovered already in 80's~\cite{Jacak:1983iz,Hahn:1986mb,Jacak:1987zz} the unstable nuclei play an important role in temperature extraction from the heavy ion collisions. Recently it was pointed out that at lower collision energies the feeddown contributions of unstable $A=4$ and $A=5$ nuclei to the
final yields of protons, deuterons, tritons, $^3$He, and $^4$He is significant and, for the latter three, may account for as much as 70\% of the final yield~\cite{Vovchenko:2020dmv}.
The same mechanism as presented in the latter paper is employed here, extending the particle list by 25 unstable nuclei that decay after freeze-out and feed into the final yields of stable light nuclei and free nucleons.
The contributions of these light nuclei improve the quality of the fit, the improvement mainly attributed to a better description of the $^3$He and $^3$H yields. 
However, the overall description is still poor, with $\chi^2/N_{\rm df}\approx 12$. 

These results suggest that either the thermal model description of light nuclei at the chemical freeze-out needs to be significantly improved or that thermal nuclei production is not the correct mechanism in the few GeV collision energy regime.

Another observation from the fits including the light nuclei is that the description of strange particles is significantly worse than in the no clusters scenario.
In general, the fits in all the studied scenarios require values of $\gamma_S$ significantly smaller than unity, meaning that the strangeness is significantly undersaturated relative to chemical equilibrium in the grand canonical ensemble.
More involved modeling of strangeness 
may thus be warranted.
In particular, the effect of canonical strangeness suppression may be particularly relevant in this energy range~\cite{Hamieh:2000tk}.

\begin{figure}[h!]
    \centering
    \includegraphics[width=0.49\textwidth]{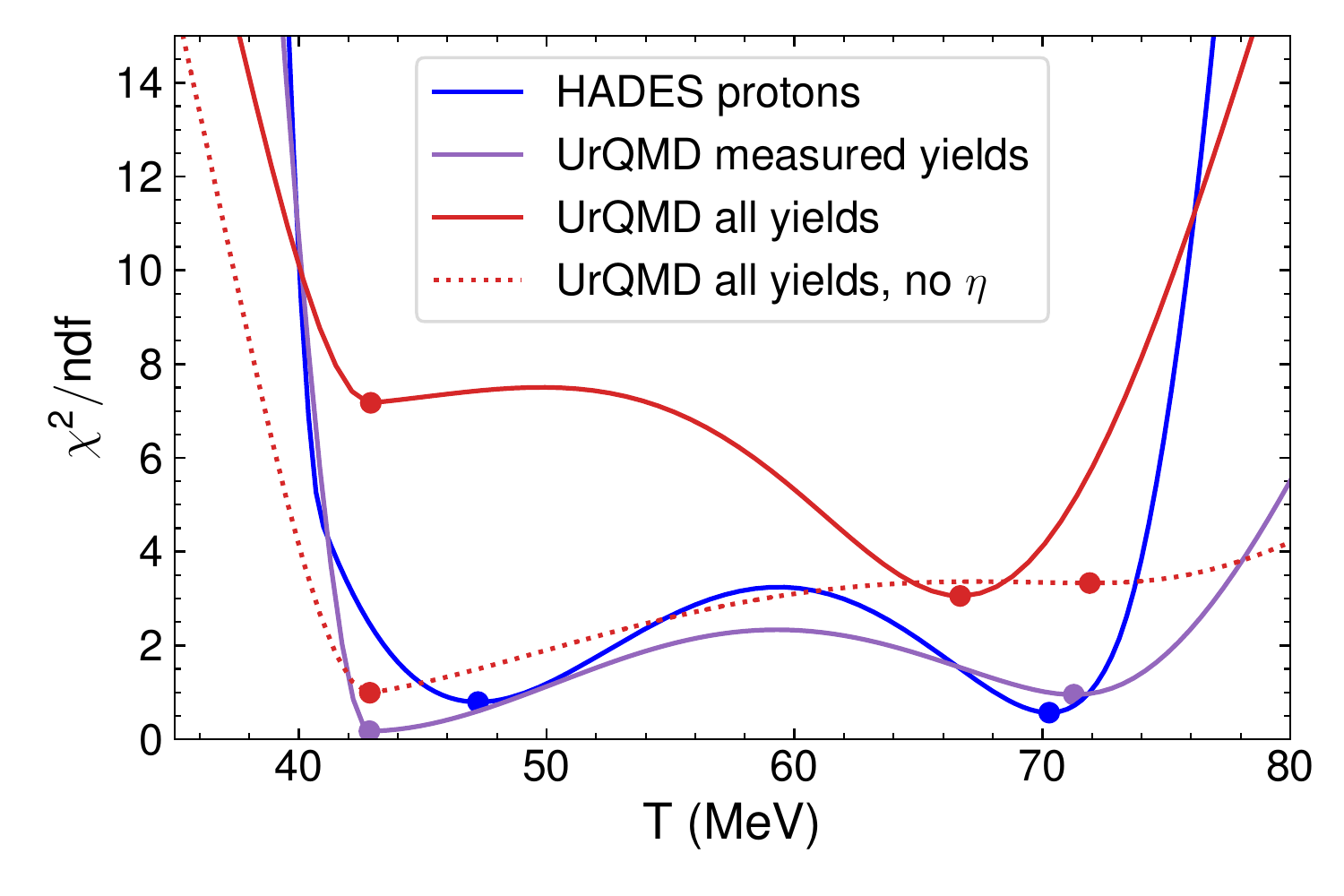}
    \caption{$\chi^2$ profiles of the fits to the experimental data and to the predicted by the UrQMD model. ``UrQMD all yields'' denotes thermal fit to all stable particle yields, ``UrQMD measured yields'' denotes fit to the UrQMD predictions for hadron species that are detected by the experiment. The UrQMD data correspond to particle yields calculated in $4\pi$ 10\% most central Au+Au events at $\sqrt{s_{_{NN}}}=2.4$~GeV. 
    The UrQMD data correspond to particle yields presented in Fig.~\ref{fig:yields_urqmd}.
    }
    \label{fig:chi2_urqmd}
\end{figure}

\section{Microscopic-macroscopic analysis of the chemical evolution with the UrQMD transport model}
\label{sec:UrQMD}

\subsection{UrQMD model setup}

In order to understand the characteristics and dependencies of the fit results from microscopic point of view, a transport model will be used to simulate heavy ion collision data. This allows for the study of the apparent origin and systematic dependencies of the two minima.
Here the heavy ion collisions are simulated with the microscopic transport model UrQMD~\cite{Bass:1998ca,Bleicher:1999xi}, where the yields of all stable hadrons can be extracted at any time step during the evolution.
To simulate the most central Au+Au collisions at \snnh 
the UrQMD model is used with a restriction $b<4.7$~fm on the impact parameter, corresponding to the 10\% most central events. All hadron yields are evaluated assuming the full detector acceptance, and include feeddown from resonance decays.
The spectator nucleons are not included in the final yields. 
In this analysis, no mechanism for light nuclei production is incorporated, hence the UrQMD analysis here corresponds to the ``no clusters'' setup from the previous section.

The UrQMD model used here is based on version 3.4, which is extended here to include an up-to-date set of resonance branching ratios. 
This extension is essential for a proper description of the sub-threshold strange particle production~\cite{Graef:2014mra,Steinheimer:2015sha}. 
Nuclear interactions play an important role at the SIS18 energy of the HADES experiment.
Therefore, our simulations incorporate the
density-dependent nuclear Skyrme potentials. 
This description gives good results for flow-observables~\cite{Hillmann:2018nmd} as well as for the space-time evolution of the density~\cite{Seck:2020qbx} at the energies available at SIS18.
Figure~\ref{fig:yields_urqmd} shows that UrQMD provides a decent description of particle yields measured by the HADES collaboration without invoking any additional free parameters.

\subsection{Thermal fitting of the UrQMD yields}

The UrQMD hadron yields listed in Fig.~\ref{fig:yields_urqmd} can be used 
in a thermal model analysis, in particular to investigate the possibility of the double minimum structure discussed in the previous section. 
Since the errors of the simulated yields are purely statistical and can be made arbitrarily small, 
for the thermal fitting procedure a 10\% relative systematic error is added, to make it comparable in magnitude to the experimental uncertainties.
In this way the thermal fitting procedure for the UrQMD yields is consistent with the procedure applied to experimental data in the previous section, thus direct comparisons can be made.

Two possibilities for the set of hadrons included in the fitting procedure are taken:
\begin{enumerate}[label=(\alph*)]
\item The fit is done only to the hadron yields which are present in the HADES collaboration data ($\pi^{+,-},~{\rm K}^{+-},~p,~\Lambda$).
\item Yields of all long-lived hadrons from UrQMD are included in the fit (see Fig.~\ref{fig:yields_urqmd} for the complete list). Here are considered separately two options regarding whether the yield of $\eta$ meson should be included or not.
\end{enumerate}
The $\chi^2$-profiles of these fits are shown in Fig.~\ref{fig:chi2_urqmd}, where they are also compared with the $\chi^2$-profile of the fit performed to the HADES data. 
The fit to UrQMD data that includes only the experimentally measured set of yields shows the two minima structure.
This is similar to the finding of the previous section using the HADES data, even the locations of both minima are similar.

However, when all the long-lived hadrons, including the $\eta$-meson, are included in the fit to the UrQMD predictions, the low temperature minimum becomes rather shallow and then the high temperature fit provides a significantly better description. 
This improved description may be considered as a direct result of the increased number of independent inputs in the fit, which constrains the fit parameters better. 
The fit quality decreases slightly, indicating tension of the thermal model with UrQMD.
This may be caused by an incomplete chemical equilibrium as well as canonical suppression that can be present in UrQMD.

It is found that the $\eta$ meson plays a particular role in the analysis.
On the one hand, the $\eta$ meson is a comparatively long-lived particle, such that it decays outside the fireball and thus treated as stable particle in UrQMD.
On the other hand, its lifetime is sufficiently short such that it decays before reaching the detector.
One can thus argue whether its yield should be included in the thermal fit or not.
If the $\eta$ meson is omitted from the fit, the $\chi^2$ profile changes such that the low temperature minimum provides a better description than the shallow, high temperature minimum.
This suggests that the inclusion of only selected additional particle multiplicities will not necessarily allow for a better distinction between the two $\chi^2$ minima. 
If the best fit depends strongly on the choice of included hadron multiplicities, these results point to the possibility that the system may  have not undergone a universal chemical freeze-out.

\begin{figure}[t]
    \centering
    \includegraphics[width=0.49\textwidth]{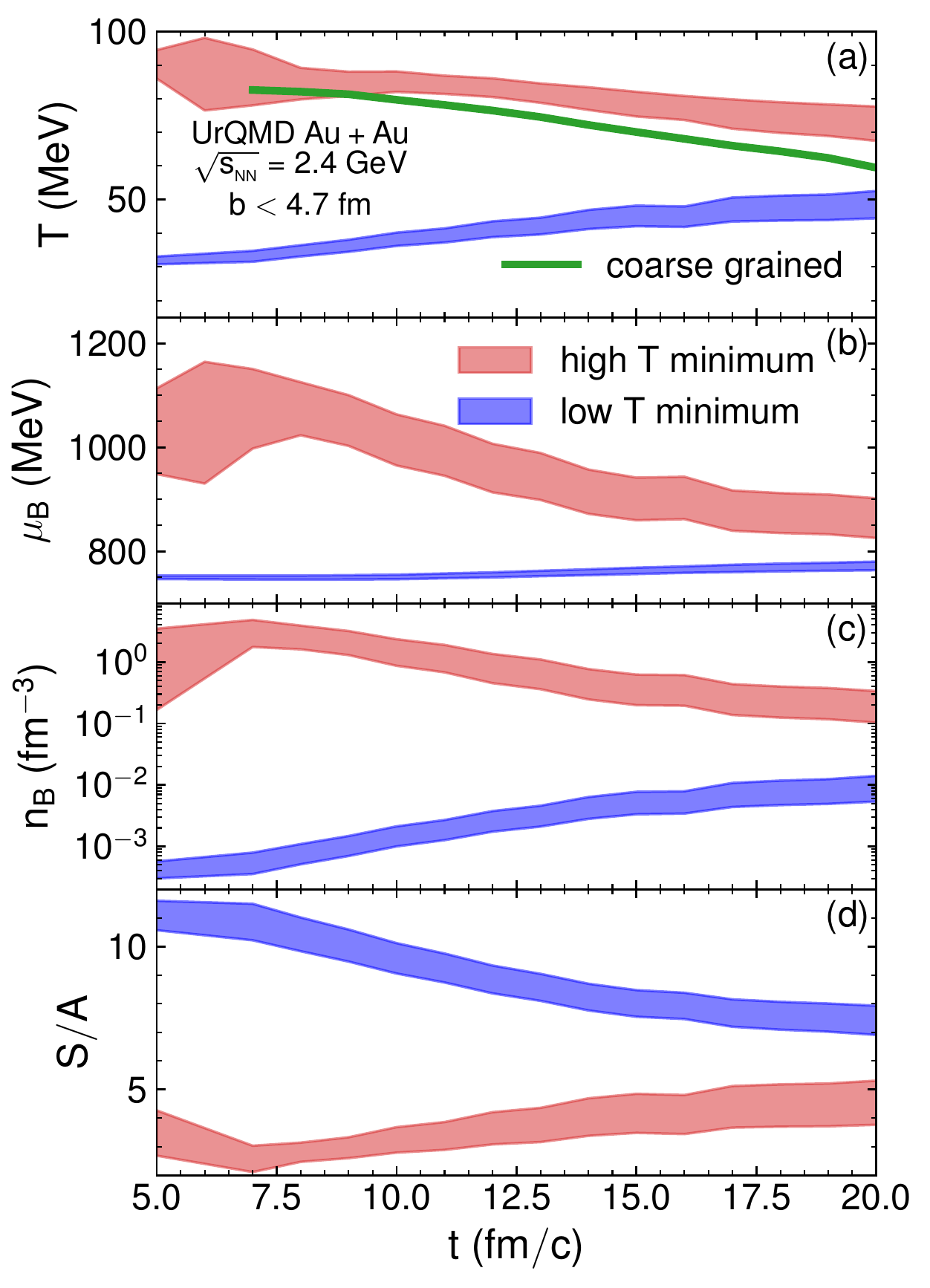}
    \caption{Time evolution of temperature $T$ {\bf (a)}, baryon chemical potential $\mu_B$ {\bf (b)}, baryon density $n_B$ {\bf (c)}, and entropy per baryon $S/A$ {\bf (d)} of chemical freeze-out extracted by thermal fits to the time dependent stable particle yields obtained with UrQMD simulations of central Au+Au collisions at \snnh. The thermal analysis presents double minima structure through the whole evolution. The low temperature minimum illustrates unphysical heating and compression (increase in the chemical potential).
    }
    \label{fig:T-mu_time}
\end{figure}

\begin{figure}[h!]
    \centering
    \includegraphics[width=0.49\textwidth]{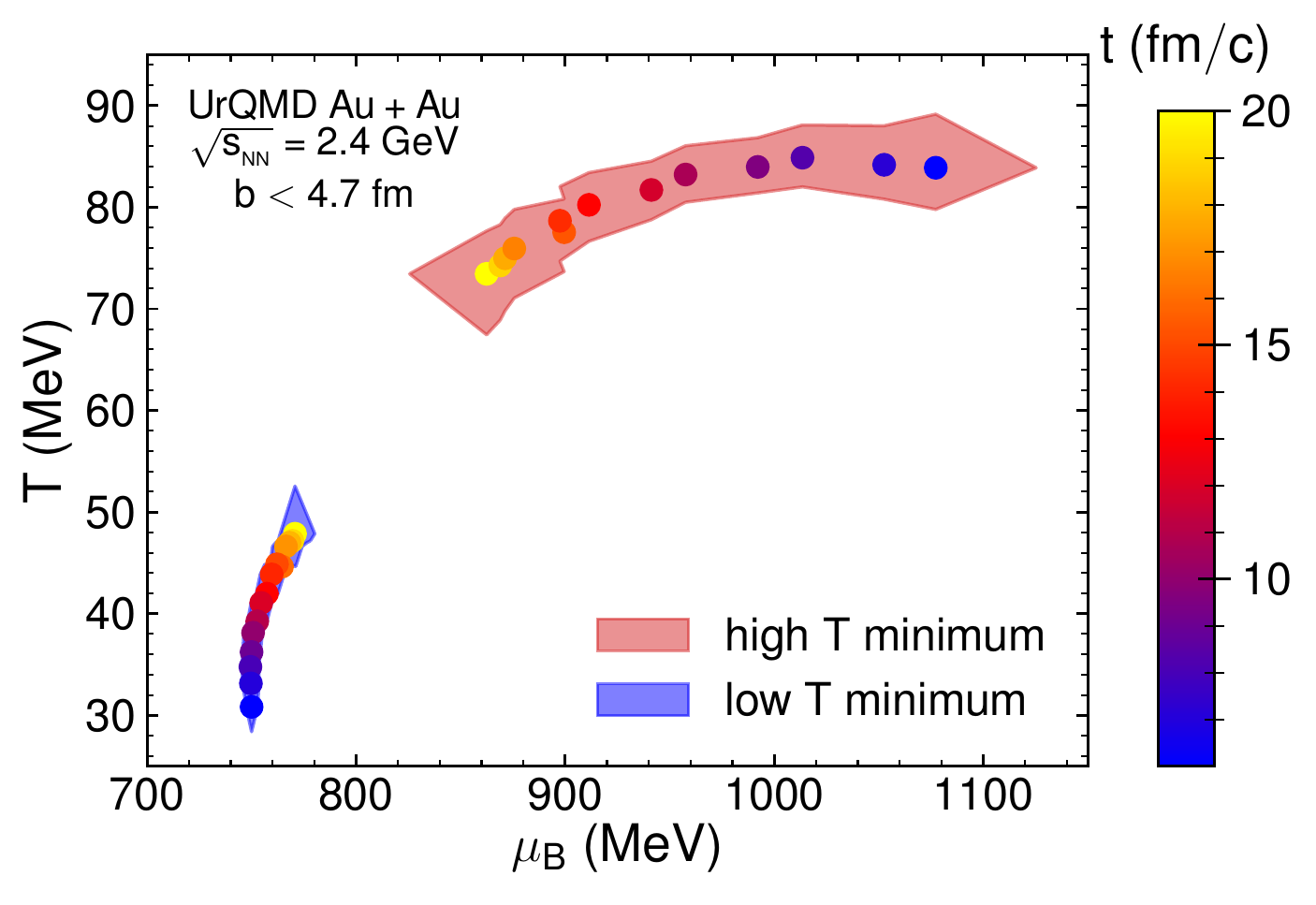}
    \caption{Time evolution of the two $\chi^2$ minima in the chemical potential -- temperature $\mu_B$--$T$ plane. The two solutions of the thermal fit evolve through two distinctly different regions of the   $\mu_B$--$T$ plane. The dots illustrate locations of the fit  at a given time, illustrated by the color. The high temperature minimum resembles a reasonable evolution of the fireball (cooling and expansion), while the lower minimum evolves unphysically with increase of the temperature and chemical potential.}
    \label{fig:muT_time}
\end{figure}

\subsection{Time evolution of the solutions}

To get a better physical understanding of the meaning of the two minima obtained in the thermal fitting the time evolution of the particle yields in UrQMD is studied following a method suggested in Ref.~\cite{Steinheimer:2016vzu}. 
To calculate the UrQMD yields corresponding to a given time moment $t$,
the transport evolution is stopped when that time moment is reached and all unstable hadron species are forced to decay. 
The calculated hadron yields are then related to the respective time $t$.
These calculated yields are then fitted with the thermal model and the extracted thermodynamic parameters are assumed to describe the colliding system at the time $t$. 
Here only the experimentally measured yields are analyzed.
The time dependence is studied in a range $t=1\mbox{-}20$~fm/$c$.

The results of the thermal fits to the time dependent yields are depicted in Fig.~\ref{fig:T-mu_time} by red (high temperature) and blue (low temperature) bands. 
The width of the bands is determined by the width of the minima in the $\chi^2/\mathrm{ndf}$ surface.  
The double minimum structure in the fit is found throughout the whole evolution. 
However, the behaviors of the thermal parameters of each minimum are opposite. 
The high temperature minimum decreases in temperature and chemical potential (the red band in figure~\ref{fig:T-mu_time}),
this solution coincides with a ``classical'' picture of the fireball evolution in a heavy ion collision: temperature and density (chemical potential) decrease with time as a result of the fireball cooling during the expansion.
The low temperature minimum on the other hand shows as a function of time a strong decrease of the total entropy per baryon, $\Delta S/A (t)< 0$. At first sight, this seems to violate the second law of thermodynamics which would indeed be a major discovery. However, be reminded that this results from global instantaneous chemical equilibrium fits to a microscopic spatio-temporal non-equilibrium model, which actually respects $\Delta S/A (t)> 0$ throughout. 

The results for the time dependence of the temperature are compared with one obtained in a different way, namely via a coarse-graining approach as in Ref.~\cite{Seck:2020qbx}~(the green line in Fig.~\ref{fig:T-mu_time}). 
Only the temperature corresponding to the high-temperature minimum is in qualitative agreement with the coarse-grained approach, indicating that the high-temperature is the only physical solution.
Note that the agreement is not fully quantitative.
In particular the coarse grained temperature, which is extracted from the local densities by thermodynamics relations rather than multiplicities, appears always lower than the chemical temperature. 
This may be another indication that the system rapidly falls out of chemical equilibrium or may not reach it fully.

Even more important, the entropy per baryon $S/A$, which can be readily calculated by the HRG, has very different behavior for the two minima as shown in  Fig.~\ref{fig:T-mu_time} (d). 
The entropy along the high temperature minimum has a slight increase, its values of $S/A \simeq 3-5$ being close to the values expected for this collision energy~\cite{Motornenko:2019arp}. 
On the other hand, the entropy per baryon of the low temperature minimum behaves abnormally, showing a decrease with time. The values of $S/A \sim 7-11$ at low temperature appear to be too high for this collision energy.
This pathological behavior of the entropy indicates that the low temperature minimum is likely unphysical.

Finally, the corresponding trajectories of the two solutions in the $\mu_B\mbox{-}T$ plane are shown in Fig.~\ref{fig:muT_time} where the respective time is indicated by color. The high temperature represents a trajectory similar to an isentropic expansion~\cite{Motornenko:2019arp} for that specific beam energy, while the low temperature evolution resembles a compression with simultaneous heating and loss of entropy.

\section{Summary}

The hadron yields measured by the HADES collaboration at SIS18 energies~\cite{Harabasz:2020sei} reveal the drawback of the thermal model approach which shows ambiguous solutions at this collision energy.
This ambiguity is reflected in the existence of two degenerate solutions of the thermal fit to the hadron yields measured in most central Au+Au collisions at \snnh by the HADES collaboration.
While the final hadron multiplicities are almost identical for both statistical model descriptions, the chemical composition and thermodynamic properties show clear differences. The high temperature solution shows a significant contribution of resonance decays to the pion yield, while the low temperature minimum  corresponds to a system which consists mainly from ground state hadrons. The high temperature solution is consistent with estimates of the $\Delta$ contribution \cite{Reichert:2020uxs} at SIS18 energies, indicating that the higher chemical freeze-out temperature can be confirmed experimentally.

The role of light nuclei production at the chemical freeze-out was also studied. The inclusion of these nuclear clusters in the fit significantly worsens the quality of the fit with $\chi^2/\mathrm{ndf} > 10$. 
In this case only one $\chi^2$ minimum is present with $T\approx 70$ MeV, consistent with the high-temperature minimum obtained without the inclusion of light nuclei. 
The fit improves if the additional feeddown from decays of unstable nuclei is included although the overall data description remains unsatisfactory.

It was found that the particle multiplicities calculated by the UrQMD transport model exhibit the same feature, i.e. can be described by the thermal model in a twofold way, with similar sets of thermodynamical parameters. The detailed study of the time evolution of same multiplicities in the UrQMD model suggests that only one of these solutions, namely the high temperature solution, shows physically reasonable behavior. 
The low temperature solution, on the other hand, behaves unphysically, with entropy that decreases with time, accompanied by increasing temperature and density.
It is concluded that only the temperature of $T\approx70$ MeV may be considered as chemical freeze-out temperature for most central \snnh Au+Au collisions. 
This high temperature solution results from a hadrochemical freeze-out that is similar in outcome to the picture of high energy collisions at SPS, RHIC, and LHC. The primordial inelastic nucleon collisions swiftly create a near chemical equilibrium early hot and dense fireball state, which subsequently expands isentropically.

Additionally, it was shown that the inclusion of further, yet unmeasured, multiplicities of stable hadrons would not allow discerning these degenerate states. 
However, inclusion of unstable hadronic species like ${\rm K}^*,\,\rho^0,\,\phi,\,\Delta$~\cite{Motornenko:2019jha, Reichert:2020uxs} in the fit would seem helpful to resolve the ambiguities. The yield of the unstable $\phi$-meson is already available from the HADES collaboration~\cite {Adamczewski-Musch:2017rtf}. This data may be used in partial chemical equilibrium approach~\cite{Motornenko:2019jha} to study in more detail the freeze-out mechanism in Au+Au collisions at SIS18 energies, this, however, requires a careful treatment of the quark content of the $\phi$-meson.

\begin{acknowledgments}
We thank Manuel Lorenz for clarification about the
HADES data and valuable suggestions.
We also thank Joachim Stroth for useful comments.
The authors are thankful for the support from HIC for FAIR and HGS-HIRe for FAIR.
JS thanks the Samson AG and the BMBF through the ErUMData project for funding. 
JS and HSt thank the Walter Greiner Gesellschaft zur F\"orderung der physikalischen Grundlagenforschung e.V. for its support. 
VV acknowledges the support through the Feodor Lynen program of the Alexander von Humboldt foundation and the U.S. Department of Energy, Office of Science, Office of Nuclear Physics, under contract number 
DE-AC02-05CH11231.
HSt acknowledges the Judah M. Eisenberg Laureatus Chair at Goethe Universit\"at Frankfurt am Main.
\end{acknowledgments}

\bibliography{main}
\end{document}